\documentclass[aps,prl,superscriptaddress,reprint,showpacs,amsfonts,amsmath]{revtex4-1}
\usepackage[usenames,dvipsnames]{xcolor}
\usepackage[final]{graphicx}
\usepackage{tikz}
\newcommand{\tlname}[1]{\ensuremath{\mathit{#1}}}

\usepackage[normalem]{ulem}

\begin{document}
\title{Residual Stresses in Glasses}
\date{\today}
\def\dlr{\affiliation{Institut f\"ur Materialphysik im Weltraum,
  Deutsches Zentrum f\"ur Luft- und Raumfahrt (DLR), 51170 K\"oln,
  Germany}}
\def\ukn{\affiliation{Fachbereich Physik, Universit\"at Konstanz,
  78457 Konstanz, Germany}}
\def\zuko{\affiliation{Zukunftskolleg, Universit\"at Konstanz,
  78457 Konstanz, Germany}}
\def\udue{\affiliation{Soft Matter Laboratory, IPKM,
  Heinrich-Heine Universit\"at D\"usseldorf, Universit\"atsstr.~1,
  40225 D\"usseldorf, Germany}}
\def\uduealt{\altaffiliation[Present address: ]{Soft Matter Laboratory, IPKM,
  Heinrich-Heine Universit\"at D\"usseldorf, Universit\"atsstr.~1,
  40225 D\"usseldorf, Germany}}
\def\hzb{\affiliation{Soft Matter and Functional Materials, Helmholtz-Zentrum f\"ur Materialien und Energie, 14109 Berlin, Germany}}
\def\hub{\affiliation{Institut f\"ur Physik, Humboldt-Universit\"at zu Berlin, 10099 Berlin, Germany}}
\def\umz{\affiliation{Institut f\"ur Physik,
  Johannes-Gutenberg-Universit\"at Mainz, Staudinger Weg 7, 55099 Mainz,
  Germany}}
\def\iesl{\affiliation{IESL-FORTH and Department of Materials Science
  and Technology, University of Crete, Heraklion 71110, Crete,
  Greece}}
\def\ufralt{\altaffiliation[Present address: ]{Department of Physics, University of Fribourg,
  CH-1700 Fribourg, Switzerland}}
\def\usaalt{\altaffiliation[Present address: ]{Institute for Theoretical Physics IV, Universit\"at Stuttgart, Pfaffenwaldring 57, 70569 Stuttgart, and Max-Planck-Institut f\"ur Intelligente Systeme, Heisenbergstra\ss{}e 3, 70569 Stuttgart, Germany}}
\def\fhalt{\altaffiliation[Present address: ]{Fraunhofer-Institut f\"ur Techno- und Wirtschaftsmathematik, Abteilung Str\"omungs- und Materialsimulation, Fraunhofer-Platz 1, 67663 Kaiserslautern, Germany}}
\author{M.~Ballauff}\hzb\hub
\author{J.~M.~Brader}\ufralt\ukn
\author{S.~U.~Egelhaaf}\udue
\author{M.~Fuchs}\ukn
\author{J.~Horbach}\uduealt\dlr
\author{N.~Koumakis}\iesl
\author{M.~Kr\"uger}\usaalt\ukn
\author{M.~Laurati}\udue
\author{K.~J.~Mutch}\udue
\author{G.~Petekidis}\iesl
\author{M.~Siebenb\"urger}\hzb\hub
\author{Th.~Voigtmann}\ukn\dlr\zuko
\author{J.~Zausch}\fhalt\umz

\begin{abstract}
The history dependence of the glasses formed from flow-melted steady states
by a sudden cessation of the shear rate $\dot\gamma$ is studied in colloidal
suspensions, by molecular dynamics simulations, and mode-coupling theory.
In an ideal glass, stresses relax only partially, leaving behind
a finite persistent residual stress. For intermediate times,
relaxation curves scale
as a function of $\dot\gamma t$, even though no flow is present.
The macroscopic stress evolution
is connected to a length scale of residual liquefaction displayed by
microscopic mean-squared displacements.
The theory describes this history dependence of glasses
sharing the same thermodynamic
state variables, but differing static properties.
\end{abstract}

\pacs{%
  64.70.P- % glass transitions of specific systems
  83.50.-v % rheology - deformation and flow
}

\maketitle

Materials are often produced by solidification from the melt,
involving nonequilibrium quenches. This imprints a history-dependent
microstructure that strongly affects macroscopic
material properties. One example is residual stresses
\cite{Withers2007,Reiter2005}:
%\cite{Withers1,*Withers2}:
if particle configurations cannot fully relax to equilibrium,
some of the stresses, arising in the presence of flow in the melt,
persist in the solid.

Small glass droplets (known as Prince Rupert's drops or Dutch tears
since the 17th century) vividly display
the effects of residual stresses \cite{Brodsley1986}:
they withstand the blow of a hammer onto their
main body, but explode when the slightest damage is inflicted upon their tail
(releasing the frozen-in stress network).
Today, safety glass and ``Gorilla glass''
covers for smartphones are deliberately pre-stressed during
production to strengthen them.
A theoretical understanding of residual stresses and their microscopic
origins is however still not achieved.

We seek to understand generic mechanisms by which residual stresses arise.
A convenient starting point is to investigate
the stress relaxation $\sigma(t)$ following the cessation of shear flow
of rate $\dot\gamma$, from a well-defined non-equilibrium stationary
state (NESS).
Such ``mechanical quenches'' are ubiquitous in soft matter,
where pre-shear is applied to ``rejuvenate'' the otherwise
ill-defined glassy state \cite{Moldenaers1986,Viasnoff2002,Negi2010,Yin2008}.
For these systems,
the soft-glassy rheology model (SGR) \cite{SGR} predicts asymptotic power laws
that imply the relaxation of stresses to zero \cite{Cates2004}.
In the following, we will reserve the term residual stress
to describe a finite, persistent stress remaining in the (ideal) glass
even at arbitrarily large times after the cessation of flow.
%Such stresses
%are clearly nonequilibrium effects that cannot be described by standard
%linear-response theory.

In addition to macroscopic rheology, we investigate
the evolution of the microscopic dynamics as
characterized by the waiting-time dependent mean-squared displacements (MSD).
The latter reveal the dynamical shrinkage of shear-fluidized regions
after cessation, and phenomena akin to,
yet different from the intensely studied aging dynamics after
thermal quenches
%(following seminal work on polymers \cite{Kovacs,Struik}).
\cite{Struik,Barrat2003}.

Experiments on a variety of colloidal suspensions, together with
molecular-dynamics (MD)
simulations, provide a coherent qualitative picture that
can be rationalized by
mode-coupling theory of the
glass transition (MCT) \cite{Goetze2009}
within the
integration-through-transients (ITT) formalism \cite{Fuchs2002}.
The theory in particular predicts the existence of a residual stress in the
glass, the magnitude of which depends on the history of shear.
We expect the same
mechanisms to be generically valid for many -- colloidal as well
as molecular -- glass formers.

The experiments, theory, and simulation are described in detail in the
Supplementary Information (SI) \cite{si} and summarized here.
We perform rheology on
colloidal poly\-styrene-PNIPAM core-shell particles in
aqueous solution (PP) \cite{Dingenouts1998,Siebenbuerger2009},
and on PMMA particles in different solvents (HS)
\cite{Koumakis.2008,Koumakis.2012,Zausch2008}.
These are well studied, nearly hard-sphere glass formers \cite{Koumakis.2012b},
but differ in particle properties like softness and polydispersity.
The average size is
$R\approx90\,\text{nm}$ (PP) and $R\approx267\,\text{nm}$ (HS) for macroscopic
rheology, and $R\approx770\,\text{nm}$ (HS) for confocal microscopy.
Density is expressed
as a dimensionless packing fraction $\varphi$; the glass
transition occurs in the two systems at
$\varphi_c\approx0.64$ (PP) and $\varphi_c\approx0.59$ (HS).
In the PP system, the effective packing fraction is sensitively tuned
through temperature.

MD simulations using a dissipative-particle-dynamics thermostat
were performed on a binary glass-forming Yukawa mixture
as outlined in Refs.~\cite{Zausch2008,Zausch2009}. The unit of length
was chosen as the small-particle radius $R$.
The glass transition temperature of this system is $T_c\approx0.14$.

ITT-MCT was evaluated numerically \cite{Fuchs2009,Brader2009,
Voigtmann2012} with the Percus-Yevick model of the equilibrium
hard-sphere structure (giving $\varphi_c\approx0.516$)
and an isotropic approximation to
spatial integrals. This model (ISHSS) has been used
together with our MD system and confocal microscopy
to study the evolution from equilibrium to the NESS
\cite{Zausch2008,Laurati2012}.
%
%ITT-MCT works with transient correlation functions
%formed with the equilibrium ensemble average.
Mean-squared displacements are calculated using a
schematic model based on Refs.~\cite{Krueger2009,Krueger2010}.
%We use an extension
%to waiting-time dependent averages on the level
%of schematic models \cite{Krueger2009,Krueger2010}.
%%, adapted to the case of shear cessation \cite{si}.

\begin{figure}
\includegraphics[width=\linewidth]{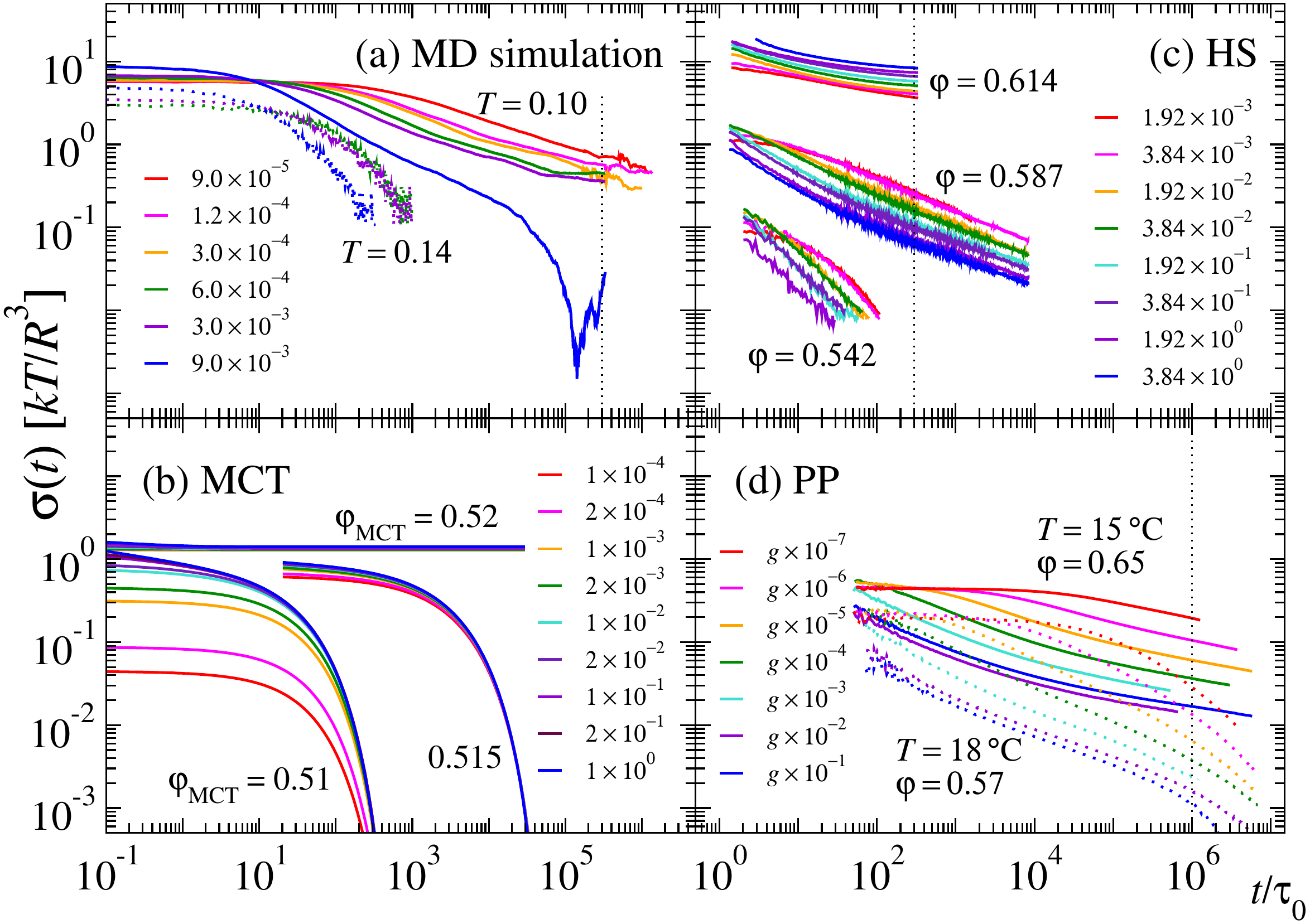}
\caption{\label{fig:sigma}
  Stress decay $\sigma(t)$ after cessation of steady shear
  (time in units of the microscopic relaxation scale $\tau_0$),
  for various shear rates $\dot\gamma\tau_0=\tlname{Pe}_0$ (increasing from
  red to blue) and control parameters, as labeled.
  Dotted lines indicate times used to extract residual stresses
  in Fig.~\ref{fig:flowcurve}.
  (a) MD simulation: $T=0.14$ in the liquid, $T=0.1$ in
  the glass.
  (b) Isotropic hard-sphere model of ITT-MCT, $\varphi_\text{MCT}=0.51$,
  $0.515$
  (liquid), and $0.52$ (glass).
  (c) HS colloidal suspension: $\varphi=0.542$, $0.587$ (liquid), and $0.614$
  (glass).
  (d) PS-PNIPAM particles: $T=18\,^\circ\text{C}$ ($\varphi\approx0.57$, liquid,
  $\dot\gamma$ as labeled with $g=3.4$)
  and $T=15\,^\circ\text{C}$ ($\varphi\approx0.65$, glass, $g=4.0$).
}
\end{figure}

Figure~\ref{fig:sigma} shows the transient decay of the shear stress $\sigma(t)$
measured in rheology and computer simulation, and calculated within
ITT-MCT. For each system, curves for various shear rates $\dot\gamma$, and
thermodynamic control variables above and below the glass transition
are shown. Stresses are reported in entropic units,
$kT/R^3\approx0.21\,\text{Pa}$ (HS), $5.33\,\text{Pa}$ and $5.82\,\text{Pa}$
(PP at $T=15\,^\circ\text{C}$ and $18\,^\circ\text{C}$).
Values measured in the
two colloidal suspensions differ by a factor $25$ in their absolute value,
consistent with their size difference.
Such scale differences do not change
the qualitative rheology of dense liquids
where structural relaxation governs the dynamics \cite{metal_overshoot}.
The shear rate $\dot\gamma$ is switched off at $t=0$ after all systems
have reached a well-defined NESS, imposing a constant strain for all $t>0$.
Times are reported relative to the
scale of single-particle motion; related to free diffusion $D_0$,
$\tau_0=R^2/D_0\approx0.3\,\text{s}$ (HS),
$4.0\,\text{ms}$ and $3.4\,\text{ms}$ (PP), or to ballistic motion
and the potential energy scale $\epsilon$,
$\tau_0=\sqrt{4mR^2/\epsilon}$ (MD). (The different forms of
short-time motion may result in a shift of the relevant $t/\tau_0$ when
comparing features of the long-time dynamics \cite{Gleim1998,Voigtmann2004b}.)
Curves start
from the corresponding steady-state value $\sigma(t\!=\!0)
=\sigma_\text{ss}(\dot\gamma)$ (separately measured),
which is a nonlinear function of the
shear rate (called the flow curve).

Stresses in dense liquids are dominated by the structural contribution,
hence slow glassy dynamics governs their decay at long times. In the
fluid, $\sigma(t)$ relaxes to zero on the structural-relaxation time scale
$\tau$. Approaching the glass transition, $\tau$ grows beyond the experimental
window, and intermediate plateaus develop in the stress relaxation.
In the MCT idealization, $\tau$ diverges as
permanent local caging of particles prevents the full relaxation of
density fluctuations.
The previously shear-molten ideal glass is then characterized
by a non-relaxing persistent residual stress $\sigma_\infty(\dot\gamma)
=\lim_{t\to\infty}\sigma(t)>0$.
Its shear-rate dependence highlights the nonequilibrium nature
of the glassy state attained after imposing zero-flow conditions:
different glasses exist with the same thermodynamic control
parameters, but different history-dependent frozen-in properties.

In simulation and experiment, the residual-stress plateaus slowly decay,
possibly as a result of creep
\cite{Siebenbuerger2012,Petekidis2003,Petekidis2004},
which is precluded in the present MCT calculations. Imposing zero-stress
instead of zero-strain-rate conditions on the other hand will allow
for rejuvenation effects that may eliminate residual stresses
\cite{Petekidis2003,Petekidis2004}.

\begin{figure}
\includegraphics[width=.9\linewidth]{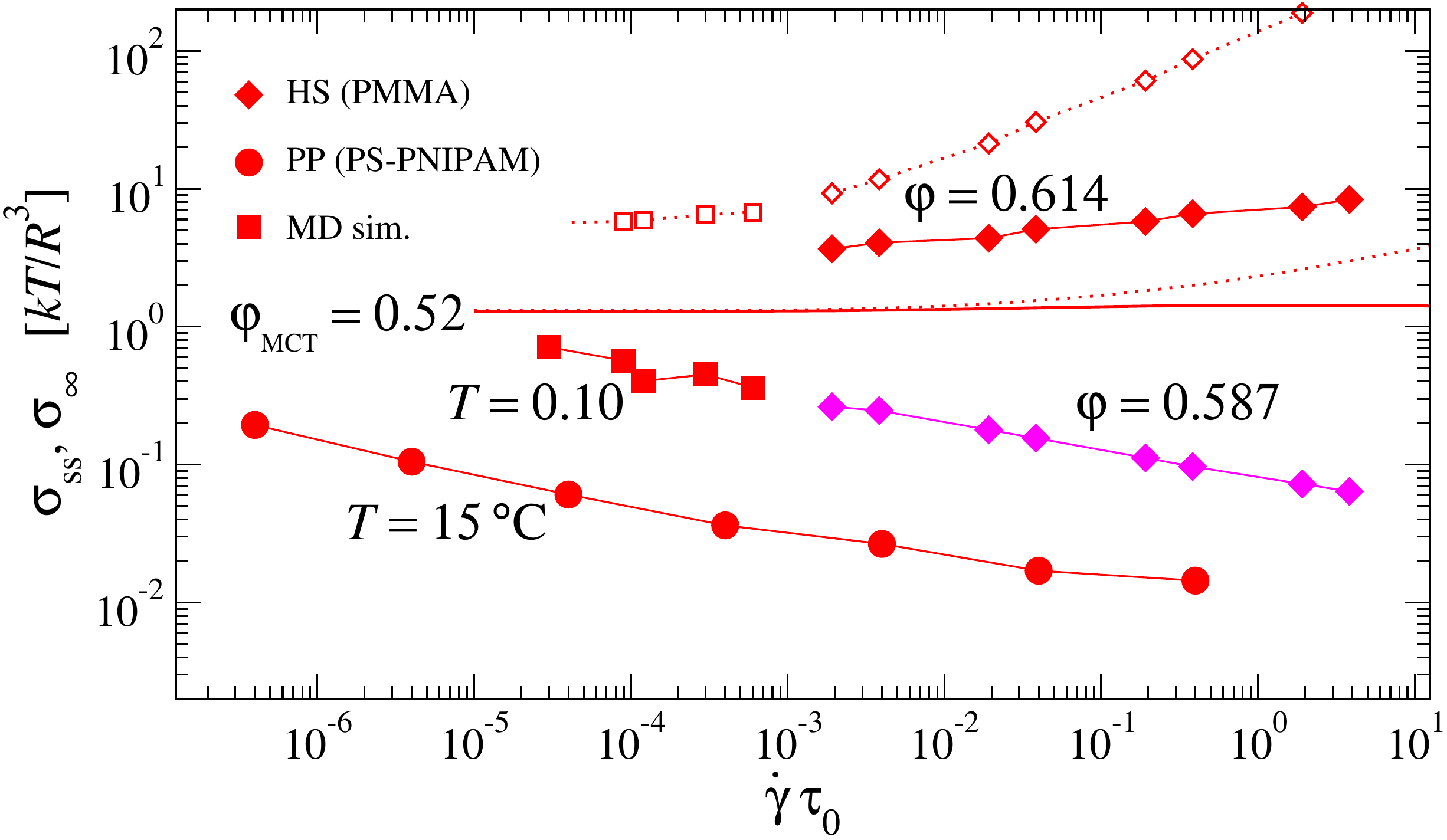}
\caption{\label{fig:flowcurve}
  Steady-state flow curve $\sigma_\text{ss}(\dot\gamma)$ (open symbols) and
  residual stress $\sigma_\infty(\dot\gamma)$ (filled)
  determined from experiment as $\sigma_\infty\approx\sigma(t_\infty)$ in
  Fig.~\ref{fig:sigma} (diamonds: HS, circles: PP) and
  MD simulation (squares). A solid (dotted) line shows
  the residual stress (flow curve) for the MCT model of Fig.~\ref{fig:sigma}.
}
\end{figure}

The residual stress $\sigma_\infty(\dot\gamma)$ as a function of the
pre-shear rate is shown in Fig.~\ref{fig:flowcurve} (filled symbols).
In experiment and simulation, comparable
values have been determined at a suitable
intermediate time $t_\infty$ (marked in Fig.~\ref{fig:sigma}).
Also shown (open symbols) are some of the flow curves
$\sigma_\text{ss}(\dot\gamma)$.
They exhibit two features typical for glass-forming
fluids: a dynamical yield stress, $\sigma_y=\lim_{\dot\gamma\to0}
\sigma_\text{ss}(\dot\gamma)>0$,
and a monotonic increase with increasing
$\dot\gamma$. The residual stresses $\sigma_\infty(\dot\gamma)$, on the
other hand, typically decrease with
increasing $\dot\gamma$ (leading to a crossing of $\sigma(t)$-versus-$t$
curves at short $t/\tau_0$, see Fig.~\ref{fig:sigma}a):
the stronger the past fluidization (and hence, structural distortions
\cite{Koumakis.2012})
of the glass, the more effective the stress relaxation.

For $\dot\gamma\to0$, the residual stress $\sigma_\infty$
approaches $\sigma_y$. In ITT-MCT, the two quantities coincide
in this limit, implying that
for arbitrarily slow flow, the glass attains a certain stress that
can never relax, even after the perturbation is removed.
To understand this, recall that the shear
stress is given by a nonlinear Green-Kubo relation,
$\sigma_{xy}(t)=\int_{-\infty}^t\dot\gamma(t')G(t,t',[\dot\gamma])\,dt'$,
where the generalized shear modulus $G(t,t',[\dot\gamma])$ is a transient
correlation function that is formed with the equilibrium
ensemble average and is only affected by external
perturbations active between its two time arguments $t'<t$ \cite{Brader2009}.
In the absence of other relaxation mechanisms,
the same physical process, shear-induced breaking of cages on a time
scale $\tau_{\dot\gamma}\sim1/\dot\gamma$,
dominates the history integrals determining
$\sigma_y$ and $\sigma_\infty$,
%In the
%limit $\dot\gamma\to0$, the different contributions at short $t-t'$
%(flowing vs.\ quiescent) become irrelevant,
and $\sigma_\infty\approx
\sigma_y>0$ results.
The ISHSS model of ITT-MCT predicts a slight increase of $\sigma_\infty$
with increasing $\dot\gamma$, which is only seen in the HS experiment
well in the glass.

\begin{figure}
\includegraphics[width=.95\linewidth]{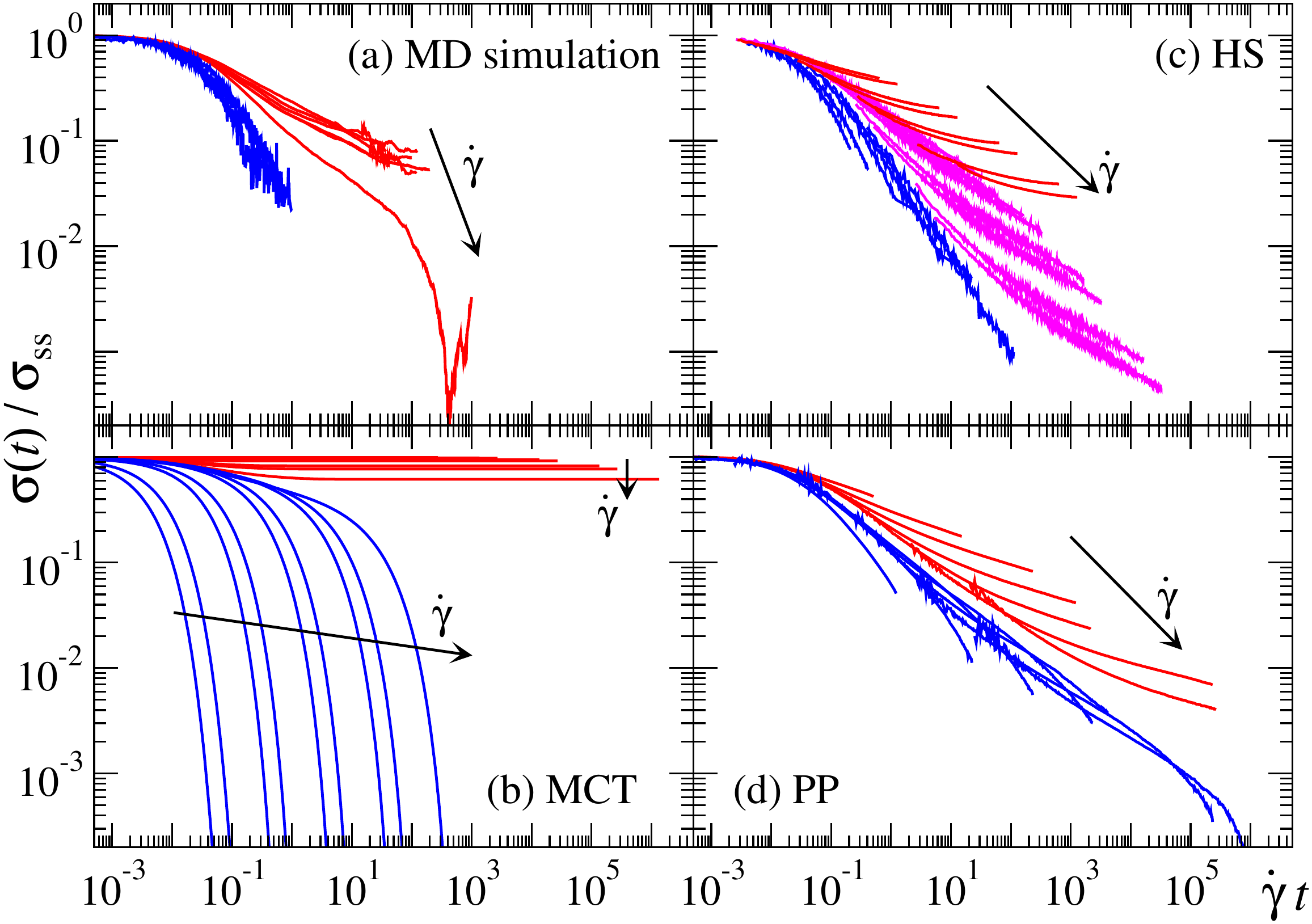}
\caption{\label{fig:sigma-scaled}
  As Fig.~\ref{fig:sigma} but $\sigma(t)/\sigma_\text{ss}$ as a function
  of $\dot\gamma t$. Glass (liquid) states are shown in red (blue).
}
\end{figure}

The stress relaxations reveal remarkable scaling behavior close to the glass transition.
Normalizing stresses by their flow-curve value, and rescaling time with
the initial shear rate, the relaxation curves fall into two classes
as shown in Fig.~\ref{fig:sigma-scaled}. Two distinct decay patterns --
one for the liquid, one for the glass -- emerge that provide a clear
indicator to locate the glass transition through a series of
shear-cessation experiments.
The dependence on ``pseudo-strain'' $\dot\gamma t$, as if
flow persisted, is remarkable since the equations of
motion at $t>0$ contain no reference to the past perturbation.
The scaling hence can be thought of as the slow decay of the NESS-contributions
to the distribution function over which dynamical quantities are
averaged. In contrast, the final decay of $\sigma(t)$
to zero in the liquid does not scale with $\dot\gamma$, as it is governed
by the equilibrium relaxation time.

The SGR predicts the observed
scaling with $\dot\gamma t$ based on aging phenomena \cite{Cates2004}.
%together with the crossing of $\sigma(t)$-versus-$t$ curves,
%which was verified in experiment \cite{Negi2010,Yin2008}.
Specifically, one obtains an asymptotic power-law
decay to zero,
$\sigma(t)/\sigma_\text{ss}\sim(\dot\gamma t)^{-x}$ for $\dot\gamma t\gg1$,
where $x$ is a temperature-like parameter.
A finite residual stress $\sigma_\infty$ is not
predicted by the SGR.

\begin{figure}
\includegraphics[width=.95\linewidth]{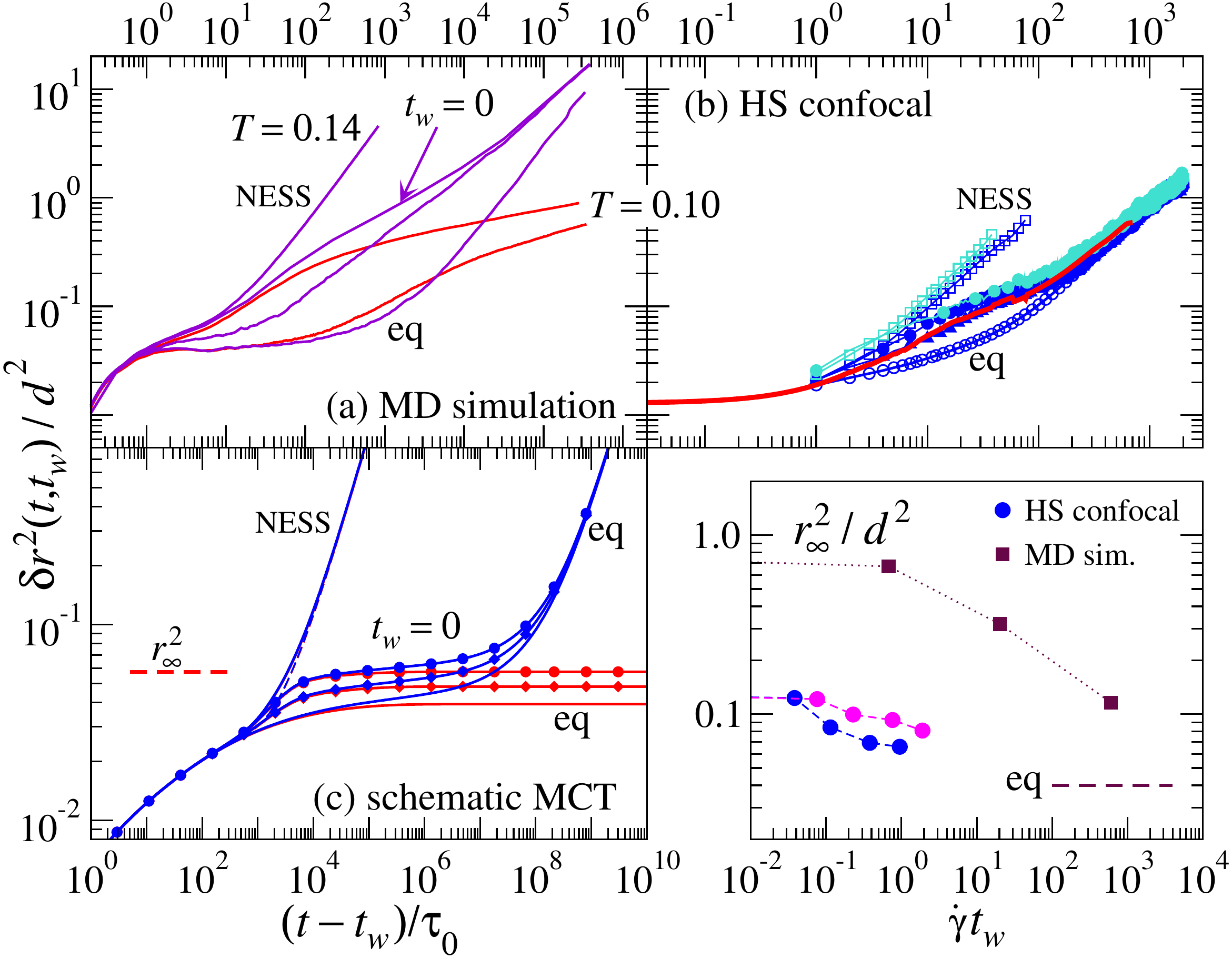}
\caption{\label{fig:msd}
  Mean-squared displacement $\delta r^2(t,t_w)$ starting at a waiting time
  $t_w$ after cessation of steady shear, in units of particle diameters.
  (a) MD simulation (large particles):
  for $T=0.14$, $\dot\gamma t_w=0$ and $0.675$
  for $\dot\gamma\tau_0=3\times10^{-3}$
  together with the NESS and equilibrium result (violet). For $T=0.1$, $t_w=0$
  curves for $\dot\gamma\tau_0=3\times10^{-3}$ and $3\times10^{-5}$
  (red; right to left).
  (b) Confocal microscopy on hard-sphere PMMA colloids, at
  $\varphi=0.56$. Curves for NESS and equilibrium together with
  $t_w=0$ for $\dot\gamma\tau_0=0.0038/\text{s}$ and $0.0076/\text{s}$
  ($\tlname{Pe}_0\approx0.0097$ and $0.019$).
  For the smaller shear rate,
  $t_w=4\,\text{s}$ is also shown. The thick red line is a theory
  fit to this curve.
  (c) Schematic ITT-MCT model, see text, for two state points
  $\varepsilon=\pm10^{-4}$ (glass/liquid) with
  $\dot\gamma\tau_0=10^{-5}$, and $t_w=0$ and $\dot\gamma\,t_w=0.01$.
  (d) Intermediate-plateau values after switch-off, $r^2_\infty$,
  as a function of $\dot\gamma\,t_w$. A dashed line marks the equilibrium
  plateau height from MD.
}
\end{figure}

To elucidate the microscopic mechanisms at play during stress relaxation,
we turn to the mean-squared displacements (MSD) of individual particles.
%\cite{vorticity}.
This is a two-time average, $\delta r^2(t,t_w)=\langle(\vec r(t)
-\vec r(t_w))^2\rangle_{t_w}$, governed by the equilibrium dynamics but
averaged with respect to the statistical
ensemble at a waiting time $t_w>0$ after cessation of shear.
MSD in the vorticity direction, for various shear rates and state points,
and for various $t_w$, from confocal-microscopy experiments and MD simulations
are shown in Fig.~\ref{fig:msd}. The $t_w$-independent
MSD measured in the equilibrium fluid and in the
sheared NESS are shown as a reference. They exhibit an intermediate-time
plateau indicating transient caging of particles in their
nearest-neighbor shells. Long-time diffusion is enhanced in the NESS,
a non-linear response effect that is the microscopic analogue of shear
thinning \cite{Besseling2007}.
Measuring the MSD with reference to
the configuration at the cessation point, $t_w=0$, the dynamics still
follows that of the NESS, up to a time given by
$\dot\gamma t\approx0.1$. This connects to a typical strain that causes
cage-breaking
\cite{Zausch2008}.
% -- thus, $\dot\gamma$ controls the decay of the
%shear-induced perturbation of the probability distribution.
The MSD then slowly crosses over to the equilibrium
dynamics, indicating an intermediate subdiffusive plateau that
is larger than the one connected to quiescent caging.
At times $t-t_w\gtrsim\tau$, all curves become independent of
$t_w$ and collapse on the equilibrium curves.
For MD simulations in the glassy state the intermediate
plateau extends past the simulated time
window. The height of this plateau decreases with increasing $\dot\gamma$.

ITT-MCT (extended to describe waiting-time
dependent two-time
averages for small $t_w$ \cite{Krueger2009,Krueger2010,si}) qualitatively
rationalizes these MSD.
Figure~\ref{fig:msd} includes schematic-model results for both
a glassy and a liquid state. A well-defined second plateau emerges
in the cross-over from the NESS to equilibrium curves and becomes permanent
in the glass.
It can be interpreted as a second length scale beyond the quiescent
localization length, arising from the competition of
shear-induced fluidization and arrest after cessation.
Intuitively, it is a ``Brems\-weg'' (stopping distance) for individual
particles caused by the progression of the distribution function from
the perturbed state at $t_w$
to the quiescent state. But note that inertial effects
play no role in the dynamics and are, by construction, absent from the theory.

The ITT-MCT model approximates the $t_w$-dependent MSD as a combination
of the equilibrium and NESS ones.
Given this information from experiments,
the evolution of $\delta r^2(t,t_w)$ for small $t_w$ can be predicted. This
is demonstrated by the fit to the HS data shown in Fig.~\ref{fig:msd}b.

The scenario of nonequilibrium relaxation discussed here
characteristically differs
from aging dynamics in spin glasses \cite{janus,Mathieu2010} or
that following density quenches in hard spheres \cite{Puertas2010,PerezAngel2011,Di2011}.
There the structural relaxation time $\tau(t_w)$ grows with sample age,
so that correlation functions and related two-time averages depend on
$t_w$ in their long-time part, while the short-time relaxation for increasing
sample age reveals
more and more of an intrinsic, $t_w$-independent relaxation.
For the MSD, this implies a continuous shift of the long-time diffusive
asymptote (where $\delta r^2\sim t$) to longer times. In contrast,
in Fig.~\ref{fig:msd} all intermediate-$t_w$ curves show subdiffusive
transients and an approach to the same $t_w$-independent long-time diffusion
found in the quiescent system.

In conclusion, we have studied stress decays and the
microscopic dynamics of glass-forming liquids and
shear-melted glasses after the cessation of steady shear flow.
In the liquid, stresses relax to zero following long transients,
on the time scale of the quiescent
equilibrium system.
(In rheological terms, the systems exhibit ``thixotropy''.)

Finite residual stresses remain in the glass.
Their value and the initial evolution from the steady state to
the nonequilibrium quiescent solid, are governed by the pre-shear rate through
long-lived memory effects.
These memory effects cause the appearance of a microscopic supra-caging
length scale in the waiting-time dependent mean-squared displacements.

Residual stresses imply that glass is not simply
characterized by its thermodynamic control variables. Different preparation
histories result in glasses that differ subtly in their structure, and
possibly also in their response behavior, for example their elastic moduli
\cite{Moldenaers1986}. Within ITT-MCT, a pre-strain dependence of the
shear modulus has been studied in the flowing steady state \cite{Frahsa2013}.
It would be enlightening to compare this to a history dependence that arises
in the case of temperature quenches, as recently studied in computer
simulations
\cite{Ashwin2013}.

The theory describes history-dependent glass states through retarded-friction
contributions to the dynamics of density fluctuations that are modified
by the past flow. This goes beyond traditional
near-equilibrium glass transition theories (such as standard MCT),
where the relaxation of small initial perturbations induced by an external
field is related to a Kubo correlator of equilibrium fluctuations.
This connection is based on Onsager's regression hypothesis, and
it holds if the initial perturbation obeys equilibrium
linear-response theory. It is violated in the glass (unless
activated processes restore ergodicity) \cite{Williams2006},
since infinitesimally small rates shear-melt the glass \cite{Fuchs2002}.
These states relax only due to external driving,
which is the common cause of both a dynamical yield stress
$\sigma_y=\lim_{\dot\gamma\to0}\sigma_\text{ss}(\dot\gamma)>0$ (contradicting
linear response), and of a residual stress $\sigma_\infty$
(contradicting equilibrium).

The qualitative agreement among different systems and methods suggests that
our discussion applies generally to glass formers where excluded volume
effects dominate,
causing stresses to be given by the entropic scale $kT/R^3$.
This includes dense metallic glass formers, where an understanding of
frozen-in residual
stresses is important for understanding material stability.

In such systems, possibly also for colloidal systems deep in the
glass, additional effects may need to be accounted for, connected to
spatial and temporal flow heterogeneities. Then, concepts such as
shear-transformation zones (STZ) \cite{STZreview}
or stress avalanches \cite{Lemaitre2009,Chattoraj2011}
(or, in the case of granular materials, force chains \cite{Hartley2003})
can become important.
For the data presented above, we have checked that no such inhomogeneities
are detectable either in confocal microscopy or during the simulation runs.

\begin{acknowledgments}
Th.~V.\ is funded by the Helmholtz Gesellschaft (HGF, VH-NG~406),
and Zukunftskolleg der Universit\"at Konstanz; M.~K.\ by
DFG grant KR~3844/2-1;
G.~P.\ and N.~K.\ by Thales Project ``Covisco'' and
EU project ``ESMI''.
We thank for funding through
DFG FOR~1394,
project P3, and SFB-TR6,
projects A5 and A6.
\end{acknowledgments}

\bibliographystyle{apsrev4-1}
\bibliography{lit}

\end{document}

\title{Supplement for ''Residual Stresses in Glasses``}
\maketitle

For completeness, we first briefly summarize the experimental and simulational
techniques used in the paper, as documented previously
\cite{Dingenouts1998,Siebenbuerger2009,Koumakis.2008,Koumakis.2012,
Koumakis.2012b,Laurati2012,Zausch2008,Zausch2009}.
After that, we summarize the equations of ITT-MCT used to describe
both stress-relaxation curves and the waiting-time dependent mean-squared
displacements, based on Refs.~\cite{Brader2009,Krueger2009,Krueger2010}.

\section{Colloidal Rheology}

\subsection{PS-PNIPAM System}

The PS-PNiPAM colloids consist of a poly(styrene) core onto which a
thermosensitive poly(N-isopropylacrylamide) shell, cross-linked with 2.5~mol-\%
N,N-methylenbisacrylamide, is affixed. The two-step synthesis is described in
detail in \cite{Dingenouts1998}. The polydispersity of the latex particles was
determined to be 17\% \cite{Siebenbuerger2009}. The colloids are suspended in
aqueous 0.05~M KCl to screen residual charges from the synthesis and show a
temperature dependent size, which can be described below 25${}^\circ$C with a
linear relationship of the hydrodynamic radius,
$R_H = − 0.7796\,\text{nm}/{}^\circ\text{C}\, T + 102.4096\,\text{nm}$,
with $T$ as the temperature in ${}^\circ\text{C}$ \cite{Siebenbuerger2009}.
Suspensions with exactly the same particles were also used for characterizing
the steady sheared state, the linear viscoelastic behavior
\cite{Siebenbuerger2009}, FT-rheology \cite{Brader2010}, start-up of shear
\cite{Amann2013} and stress \cite{Siebenbuerger2012}.

The cessation of the suspension in this work was tested at $15{}^\circ\text{C}$
and $18{}^\circ\text{C}$ which correspond to a volume fraction
$\varphi_\text{eff} = 0.65$ and $\varphi_\text{eff} = 0.60$, taking into
account the solid content of the suspension of
8.35 wt-\% (see Ref.~\cite{Siebenbuerger2009} for a detailed description of the
effective volume fraction calculation). The rheological experiments were
performed with the stress-controlled MCR 301 from Anton Paar, using a
cone-plate geometry with a diameter of $50\,\text{mm}$ and a cone angle of
$0.017\,\text{rad}$. Here, a thin film of paraffin in combination with a
covering cap prevented evaporation. 

A two-step protocol for the cessation experiments was used: first a steady
sheared state at a given shear rate was reached by shearing at minimum the
time span of the inverse shear rate; in the second step the deformation was
set to $\gamma = 0$ and the cessation stress was recorded. The values of the
sheared steady state before the switch off are found to coincide with the flow
curves, which confirm the status of the steady sheared state before switch off. 

\subsection{PMMA System}

We have performed rheological measurements on Polymethylmethacrylate
(PMMA) particles, sterically stabilized by a thin chemically grafted layer
of poly-12-hydroxystearic acid chains and suspended in cis-trans decalin
allowing for nearly hard sphere interactions \cite{Bryant.2002}. The
particle radii were determined by light scattering to be R=267nm with a
polydispersity of approximately 6\%. Volume fractions were prepared by
progressively diluting a single batch, having the volume fraction
determined in the coexistence regime \cite{puseybook}.

Rheological measurements were conducted on an Anton-Paar MCR501 rheometer
with cone-plate geometries (25 and 50 mm diameter and 0.01 rad angle)
using a solvent trap to minimize evaporation. Constant rate and stress
relaxation tests were performed for various volume fractions 0.542, 0.587
(liquid), and 0.614 (glass) with a glass transition at 0.60 as
rheologically estimated \cite{Koumakis.2012b}.

Constant rate experiments were allowed to reach a steady state ($>100\%$
strain) before enforcing the deformation rate to zero and measuring the
stress. Although the samples were monodisperse, shear-induced
crystallization was suppressed by avoiding large amplitude oscillatory
measurements \cite{Koumakis.2008}.

\section{Confocal Microscopy}

\subsection{Samples}

We used PHSA sterically stabilized Polymethylmethacrylate (PMMA) colloidal
spheres, fluorescently labeled with nitrobenzoxadiazole (NBD). The spheres
were dispersed in a mixture of cycloheptyl bromide and cis-decalin,
approximately matching both the density and refractive index of the colloids. 
Charges, present in small amount in this solvent mixture, were screened by
adding $4$~mM tetrabutylammoniumchloride \cite{yethiraj03}, obtaining a system
which displays almost hard-sphere behaviour \cite{Poon/Weeks/Royall}. 

We determined the particle radius $R=770\pm6$~nm  and the polydispersity of
approximately 6\% using static and dynamic light scattering measurements of a
very dilute colloidal suspension ($\varphi\simeq10^{-4}$). A similar value of
$R=780$~nm was obtained by confocal microscopy. Samples for
confocal microscopy under shear, at $\varphi = 0.56 \pm 0.01$
\cite{Poon/Weeks/Royall}, were prepared from a random-close-packed (RCP) stock
solution which was obtained by sedimenting a dilute colloidal suspension in a
centrifuge. The volume fraction of the sediment ($\varphi_\text{rcp}$) was
determined as follows: A first guess for $\varphi_\text{rcp}$ was obtained
from simulations and used to dilute the sediment to a nominal
$\varphi\approx0.4$.
The experimental volume fraction of this dispersion was obtained by imaging
from the ratio of the particle volume to the mean Voronoi volume and used to
recalculate $\varphi_\text{rcp}\approx 0.68$.
%%%%% confocal

\subsection{\label{sec:confocal}Confocal Microscopy under Shear}

Shear was applied to the sample by means of a parallel plates home-built shear
cell \cite{laurati:jpcm12,petekidis02PhysicaA}. 
The plates, two glass coverslips, were coated with a layer of very polydisperse
PMMA particles \cite{ballesta08} to prevent wall slip. 

Imaging of the samples in the shear-cell was performed using a VT-Eye confocal
microscope (Visitech International) mounted on a Nikon TE2000-U inverted
microscope, using a Nikon Plan Apo VC 100$\times$ oil immersion objective. 
In order to avoid wall effects and retain a good signal-to-noise ratio, images
of $512\times512$ pixels were recorded in a plane located $30\,\text{$\mu$m}$
inside the sample. Each image corresponds to an area of
$57\times57\,\text{$\mu$m}^2$. 
A series of images was acquired at a fast sampling rate (compared to the
particle motions), starting at the moment where the application of shear was
stopped.

Particle coordinates and trajectories were obtained from images using standard
routines \cite{crocker96}. 
Typically the imaged area contains about 1200 particles.  
Mean squared displacements (MSD) in the vorticity direction,
$\langle \delta z^2(t,t_{w}) \rangle$, were calculated from trajectories as a
function of time $t$ and for different values of the waiting time
$t_{w}$, which quantifies the delay after cessation of shear. 
MSDs extracted from typically five to ten experiments were averaged after
checking the reproducibility of the measurements.

\section{Molecular Dynamics Simulations}

Molecular dynamics computer simulations have been done for a
binary mixture of charged colloids. Interactions
between the particles are modeled by a Yukawa potential,
\begin{equation}
\label{eq_pot}
u_{\alpha\beta}=
\epsilon_{\alpha \beta} d_{\alpha \beta}
\frac{\exp(- \kappa_{ \alpha \beta}(r- d_{\alpha\beta}))}{r}
\quad \alpha,\beta=\rm{A,B},
\end{equation}
truncated at a cut-off distance $r_{\rm
c}^{\alpha\beta}$, defined by $u_{\alpha\beta}(r_{\rm
c}^{\alpha\beta})=10^{-7}\,\epsilon_{\rm AA}$.  The ``particle diameters''
are set to $d\equiv d_{\rm AA}=1.0$, $d_{\rm BB}=1.2\,d$ and $d_{\rm
AB}=1.1\,d$, the energy parameters to $\epsilon\equiv\epsilon_{\rm
AA}=1.0$, $\epsilon_{\rm BB}=2.0\,\epsilon$, $\epsilon_{\rm
AB}=1.4\,\epsilon$, and the screening parameters to $\kappa_{\rm
AA}=\kappa_{\rm BB}=\kappa_{\rm AB}=6/d$. The choice of these parameters
ensures that, at the density $\varrho=0.675\,m_{\rm A}/d_{\rm
AA}^3$ considered, no problems with crystallization
or phase-separation occur, at least in the temperature range under
consideration.  The masses of the particles are set to unity, i.e.~$m=m_{\rm
A}=m_{\rm B}=1.0$.

The simulations were done for a 50:50 mixture of $N=2N_{\rm A}=2N_{\rm
B}=1600$ particles, placed in a cubic simulation box of linear size
$L=13.3\,d$. For the sheared system, we chose the $x$ direction as the
direction of shear and the $y$ and $z$ direction as the gradient and
vorticity direction, respectively.  Shear was imposed onto the system
via modified periodic boundary conditions, the so-called Lees-Edwards
boundary conditions \cite{lees72,evansbook}.
The application of
of Lees-Edwards boundary conditions leads to a linear shear profile
in the steady state regime, $v_{{\rm s},x}(y)=\dot{\gamma} (y-L/2)$ with the shear
rate $\dot{\gamma}=u_{{\rm s},x}/L$.

The system was coupled to a dissipative particle dynamics (DPD) thermostat
\cite{soddemann03}. The DPD equations
of motion are given by
\begin{equation}
\dot{{\bf r}}_i=
\frac{{\bf p}_i}{m_i},\quad
\dot{{\bf p}_i}=\sum_{j(\neq i)}
   \left[{\bf F}_{ij}+{\bf F}^{\rm D}_{ij}+{\bf F}^{\rm R}_{ij}\right]
\label{eq_dpd}
\end{equation}
with ${\bf r}_i$ and ${\bf p}_i$ the position and momenta
of a particle $i$ ($i=1,..., N$).  In Eq.~(\ref{eq_dpd})
${\bf F}_{ij}=-{\nabla}u_{ij}$ denote the conservative force between a
particle $i$ and a particle $j$ due to the interaction potential defined
by Eq.~(\ref{eq_pot}). To provide the thermostatting of the system a
dissipative force ${\bf F}^{\rm D}_{ij}$ and a random force ${\bf F}^{\rm
R}_{ij}$ are added in Eq.~(\ref{eq_dpd}).

The dissipative force is defined by \cite{espanol95}
\begin{equation}
\label{eq_dissf}
  {\bf F}^{\rm D}_{ij}=-\zeta\,w^2(r_{ij})\,
\left(\hat{{\bf r}}_{ij}\cdot {\bf v}_{ij}\right)\,\hat{{\bf r}}_{ij}  
\end{equation}
with $\zeta$ being a friction coefficient,
${\bf v}_{ij}={\bf v}_i-{\bf v}_j$ the relative velocity between particle
$i$ and $j$, $\hat{{\bf r}}_{ij}$ the unit vector of the vector ${\bf r}_{ij}=
{\bf r}_i - {\bf r}_j$, and $r_{ij}=| {\bf r}_i - {\bf r}_j |$ the
distance between particle $i$ and $j$. The function $w(r_{ij})$ is defined
by $w(r_{ij})=\sqrt{1-r_{ij}/r_c}$ for $r_{ij}<r_c^{\rm DPD}=1.25\,d$
and $w(r_{ij})=0$ otherwise. Thus, the force ${\bf F}^{\rm D}$ describes
a frictional force due to the interaction between neighboring particle
pairs. The use of the relative velocities between neighboring particles
in (\ref{eq_dissf}) is crucial to obtain the correct behaviour on hydrodynamic
scales. It ensures Galilean invariance and local momentum conservation.
For the friction coefficient
we chose the value $\zeta=12$.

The random force in (\ref{eq_dpd}) is given by ${\bf F}^{\rm
R}_{ij}=\sqrt{2k_B T\zeta}\,w(r_{ij})\,\theta_{ij}\, \hat{{\bf r}}_{ij}$
where $\theta_{ij}=\theta_{ji}$ are uniform random numbers with zero
mean and unit variance. The amplitude of the random force, $\sqrt{2k_B
T\zeta}$, is chosen in accordance with the fluctuation-dissipation
theorem.
The equations of motion were integrated by a generalized form of the
velocity Verlet algorithm proposed by Peters
\cite{peters04}.  For the time step of the integration we used $\delta
t=0.0083\,\tau$ (with the time unit $\tau=\sqrt{m d^2/\epsilon}$).

For both temperatures, $T=0.14$ and $T=0.10$, 250 independent runs were
done, each
of them over at least 40 million time steps. The shear stress is evaluated as
\begin{equation}
 \langle \sigma_{xy} \rangle =
  \frac{1}{L^3} \left\langle \sum_i 
   \Bigl[ m_i  \bar v_{i,x} v_{i,y}
         + \sum_{j>i} r_{ij,x} F_{ij, y} \Bigr] \right\rangle \; ,
\end{equation}
where $\bar v$ is the fluctuating velocity where the drift velocity
has been subtracted.

The mean-squared displacement was evaluated for the larger B particles.
Note that for the sheared system, the MSD
are anisotropic and depend on the considered Cartesian direction.
This anisotropy persists in the transient regime after shear cessation.
We restrict our discussion to the vorticity direction $\alpha=z$, but
we have checked that the displacements in the gradient direction
$\alpha=y$ give almost the same results (confirming that shear-induced
anisotropies are small on the level of two-point correlation functions).

\section{ITT-MCT}
\subsection{Transient Correlations}
In order to theoretically describe the dynamics as measured in our experiments and simulations, we use the standard model system for MCT-ITT \cite{Fuchs2009}: $N$ spherical Brownian particles of diameter $d$, with bare
diffusivity $D_0$, and interacting via internal forces ${\bf
F}_i=-\boldsymbol{\partial}_i U$, $i=1,\dots,N$, are dispersed in a
solvent
with a homogeneous but time dependent velocity profile ${\bf v}({\bf
r},t)=\kap(t)\cdot{\bf r}$. Shear enters via the shear rate tensor
$\kap(t)=\dot\gamma(t){\bf\hat x\hat y}$. Neglecting hydrodynamic
interactions, the distribution of particle positions evolves
according to the Smoluchowski equation \cite{Dhont}
\begin{equation}\label{smol}
\partial_t \Psi(t)=\Omega(t) \; \Psi(t),\hspace{0.5cm} \Omega(t)=\sum_{i}\boldsymbol{\partial}_i\cdot\left[\boldsymbol{\partial}_i-{\bf F}_i - \kap(t)\cdot {\bf r}_i\right],
\end{equation}
where $\Omega$ is the Smoluchowski operator and we have introduced
dimensionless units for length, energy and time, $d=k_BT=D_0=1$. The
Smoluchowski operator for the system without shear ($\kap=\bf 0$ after switch-off)
is denoted $\Omega_e$, the one for steady shear (before switch-off) is denoted $\Omega_s$. The distribution for the NESS (with steady shear) is denoted $\Psi_s$, and $\left\langle\dots\right\rangle^{(\dot\gamma)}$ denotes the corresponding stationary  average (i.e., with $\Psi_s$).

ITT-MCT is formulated on the basis of transient correlation functions,
formed by an equilibrium average (taking the quiescent Boltzmann distribution
function), but fully incorporating the nonlinear shear perturbation in the
time-evolution operator. Single-time averages are expressed as history
integrals over these transient correlators. For the deviatoric stress
tensor, one
obtains (see, e.g., Ref.~\cite{Brader2009})
\begin{eqnarray}
\sig(t) &=& -\int_{-\infty}^{t} \!\!\!\!\!\!dt'\!\int\!\!\!
\frac{d{\bf k}}{32\pi^3} 
\left[
\frac{\partial}{\partial t'}(\kb\!\cdot\!\boldsymbol B(t,t')\!\cdot\!\kb)
\,\kb\kb\right]\times \label{nonlinear}
\\
&\times&
\left[
\left(\!\frac{S'_k S'_{k(t,t')}}{k\,k(t,t')S^2_k}\!\right)\Phi_{{\bf k}(t,t')}^2
\right].\notag
\end{eqnarray}
Here, $\Phi_{\bf k}(t,t')$ is the transient density auto-correlation
function for a fluctuation at wave-vector $\bf k$, taking into account
the affine deformation: density fluctuations at two different times $t'$
and $t$ overlap if their wave vectors are related through an affine
advection, ${\bf k}(t,t')={\bf k}\cdot\bf E(t,t')$. Here, ${\bf E}(t,t')$ is
the deformation tensor, related to the strain-rate tensor by
$\partial\ln{\bf E}(t,t')/\partial t={\bf\kappa}(t)$. In Eq.~\eqref{nonlinear},
the desired invariance of the stress tensor under material-frame rotations
is guaranteed by the appearance of the Finger tensor,
${\bf B}={\bf E}\cdot{\bf E}^T$. Information on the particle interactions enters
entirely through the quiescent-equilibrium static structure factor $S_k$.
Equation~\eqref{nonlinear} is easily specified for the planar shear flow
considered in the present paper, by letting $\kappa_{xy}=\dot\gamma$ for
$t<0$, and zero else.

The transient density correlation functions are given by a Mori-Zwanzig-type
memory equation,
\begin{eqnarray}
\dot\Phi_{\bf q}(t,t_0)
&+& \Gamma_{\bf q}(t,t_0)\bigg(
\Phi_{\qb}(t,t_0)
\label{equom}
\\
&+&
\int_{t_0}^t dt' m_{\qb}(t,t',t_0) \dot\Phi_{\qb}(t',t_0)
\bigg) =0
\notag
\end{eqnarray}
where the overdots denote partial differentiation with respect to the first time argument.
Here the `initial decay rate' obeys
$\Gamma_{\bf q}(t,t_0)=D_0\bar{q}^2(t,t_0)/S_{\bar{q}(t,t_0)}$ with $D_0$
a bare diffusivity, and $\bar{{\bf q}}(t,t') = {\bf q}\cdot{\bf E}^{-1}(t,t')$. 
%and $\bar{{\bf q}}(t,t') = {\bf q}\cdot \exp_-[-\int_{t'}^t\kap(s)ds]$, 
%where $\exp_-[.]$ denotes a time-ordered exponential \cite{brader2}. 
Finally, MCT-ITT approximates the memory kernel $m_{\bf q}(t,t',t_0)$ by the factorized expression
\begin{eqnarray}
\label{approxmemory}
&& \hspace*{-1cm}
m_{\qb}(t,t'\!,t_0) \!\!= \!\!
\frac{\rho}{16\pi^3} \!\!\int \!\! d\kb
\frac{S_{\bar{q}(t,t_0)} S_{\bar{k}(t',t_0)} S_{\bar{p}(t',t_0)} }
{\bar{q}^2(t',t_0) \bar{q}^2(t,t_0)}\\
&\times&
V_{\qb\kb\pb}(t',t_0)\,V_{\qb\kb\pb}(t,t_0)\Phi_{\bar{\kb}(t',t_0)}(t,t')
\Phi_{\bar{\pb}(t',t_0)}(t,t'),
\notag
\end{eqnarray}
where $\pb=\qb-\kb$, and the vertex function obeys
\begin{eqnarray}
\!\!\!\!\!V_{\qb\kb\pb}(t,t_0) \!=\! \bar\qb(t,t_0)\cdot(
\bar\kb(t,t_0) c_{\bar{k}(t,t_0)} \!+
\bar\pb(t,t_0) c_{\bar{p}(t,t_0)}),
\label{vertex}
\end{eqnarray}
with $c_k = 1-1/S_k$.

Equations~\eqref{nonlinear}--\eqref{vertex} can be solved numerically,
requiring as input a model for the equilibrium static structure factor
$S_k$. For the latter, we employ the Percus-Yevick approximation for
hard spheres \cite{Hansen}. The numerical effort to solve the equations
with full wave-vector dependence is still not feasible; we therefore
resort to an isotropic approximation of the memory kernel outlined
in Ref.~\cite{Fuchs2009}.

For the $t_w$-dependent MSD, further approximations are needed, outlined
below. The numerics in this case was performed for a schematic model,
where all wave-vector dependence in Eqs.~\eqref{nonlinear}--\eqref{vertex}
was dropped, and the memory kernel replaced by a simple quadratic polynomial
in $\Phi(t,t')$, the so-called F${}_{12}$ model \cite{Brader2009}.

\subsection{Two-Time Averages}

The correlation function of particle positions after switch off of shear depends on the correlation time $t$ as well as the waiting time $t_w$ as defined in the main text. We start with the correlator $C_{\bf q}(t,t_w)$ for the single particle density $\varrho^s_{\bf q}=e^{i\vct{q}\cdot\vct{r}_s}$, which --as a special case-- contains the mean squared displacement (see below). 
Considering first the case $t_w=0$ and restricting our attention to non-advected directions with $q_x=0$, we have the exact starting point
\begin{equation}
C_{\bf q}(t,t_w=0)=\left\langle\varrho_{\bf{q}}^{s*}e^{\Omega^{\dagger}_e t}
\varrho^s_{\bf{q}}\right\rangle^{(\dot\gamma)}.
\label{corr_def}
\end{equation} 
This form can be verified by starting from the joint probability \cite{Risken} and noting that the Smoluchowski equation is local in time. The correlator in Eq.~\eqref{corr_def} follows equilibrium dynamics given by $\Omega^{\dagger}_e$ (as there is no shear field after switch-off), but information about the pre-existing steady state is nontrivially encoded in $\Psi_s$. We aim to connect $C_{\bf q}(t,t_w=0)$ to known correlation functions, and use a well known operator identity to have 
%(denoted here by $C^{SO}_{\bf q}(t)$) and the steady state correlator $C_{\bf q}(t)$
\begin{align}
C_{\bf q}(t,t_w=0)&=C_{\bf q}(t)\notag\\&-\int_0^t dt'\left\langle \varrho_{\bf q}^{s*}e^{\Omega^\dagger_s(t-t')}
\delta\Omega^\dagger e^{\Omega^\dagger_et'}\varrho_{\bf q}^{s}\right\rangle^{(\dot\gamma)}.
\label{exact}
\end{align}
$C_{\bf q}(t)$ denotes the correlation function in the NESS, and $\delta\Omega^\dagger\equiv \Omega_s^\dagger-\Omega_e^\dagger$ is the shear-part of $\Omega_s^\dagger$. Eq.~\eqref{exact} serves to formally identify the difference between steady state and $t_w=0$ correlators.  In order to make this exact result tractable we employ the projector 
$P=\varrho_{\bf q}^{s}\rangle^{(\dot\gamma)}\langle \varrho_{\bf q}^{s*}$, 
 to obtain the factorization approximation 
\begin{align}
&C_{\bf q}(t,t_w=0)\approx C_{\bf q}(t)\notag\\&-\int_0^t dt'
\left\langle \varrho_{\bf q}^{s*}e^{\Omega^\dagger_s(t-t')}
\varrho_{\bf q}^{s}\right\rangle^{(\dot\gamma)}\left\langle \varrho_{\bf q}^{s*}
\delta\Omega^\dagger e^{\Omega^\dagger_et'}\varrho_{\bf q}^{s}\right\rangle^{(\dot\gamma)}.
\label{proj1}
\end{align}
The first term in the integral is identified with the stationary correlator $C_{\bf q}(t-t')$. With $q_x=0$, and using the integration through transients formalism \cite{Fuchs2005} enables to identify the second term in the integral in Eq.(\ref{proj1}) as the waiting time derivative at zero waiting time (compare the analogous relations for the case of switch-on of shear \cite{Krueger2009,Krueger2010})
\begin{equation}
\left\langle \varrho_{\bf q}^{s*}\delta\Omega^\dagger 
e^{\Omega^\dagger_et}\varrho_{\bf q}^{s}\right\rangle^{(\dot\gamma)}
=-\left.\frac{\partial}{\partial t_w}C_{\bf q}(t,t_w)\right|_{t_w=0}. 
\label{wtd}
\end{equation}
Substituting (\ref{wtd}) into (\ref{proj1}) leads to our key result
\begin{equation}
C_{\bf q}(t,0)\approx C_{\bf q}(t)+\int_0^t dt' C_{\bf q}(t-t') 
\left.\frac{\partial}{\partial t_w}C_{\bf q}(t',t_w)\right|_{t_w=0}.
\label{eq:approx}
\end{equation}
Up to this point the only approximation is the factorization required to go from 
(\ref{exact}) to (\ref{proj1}).
It is left to approximate the waiting time derivative in Eq.~(\ref{eq:approx}). Motivated by the fact that (\ref{eq:approx}) is insensitive to the details of the approximation made for the waiting time derivative (provided that desirable limiting constraints are satisfied), we 
use the limiting values for $t_w=0$ and $t_w\to\infty$ for a rough differential quotient for the derivative with respect to $t_w$
\begin{equation}
\left.\frac{\partial}{\partial t_w}C_{\bf q}(t,t_w)\right|_{t_w=0}\approx\frac{C_{\bf q}(t,t_w\to\infty)-C_{\bf q}(t,t_w=0)}{\tau_\sigma}.
\label{wtd_2}
\end{equation}
Here we use the physically plausible assumption that the waiting time dependence relaxes on the same timescale $\tau_\sigma$ as does the stress $\sigma$.
This time scale is related to the $\beta$-relaxation time of MCT,
here denoted as $\tau_\sigma$. Note that $\tau_\sigma$ remains finite in the
glass. We can further write $C_{\bf q}(t,\infty)\approx C^{eq}_q(t)$ (the correlatior for the quiescent equilibrium system), which at least in the fluid holds exactly, and Eq.~\eqref{eq:approx} is closed {\it selfconsistently},
\begin{align}
&C_{\bf q}(t,t_w=0)\approx C_{\bf q}(t)\notag\\&+\int_0^t dt'C_{\bf q}(t-t') \frac{C^{eq}_q(t') - C_{\bf q}(t',t_w=0)}{\tau_\sigma}.
\label{eq:fin}
\end{align}
An intuitive picture arises by considering $t\gg\frac{1}{\dot\gamma}$, where $C_{\bf q}(t)=\int _0^\infty dt'C_{\bf q}(t')\delta(t)$ can be written, and a straight forward analysis of Eq.~\eqref{eq:fin} in Laplace space yields 
\begin{eqnarray}
C_{\bf q}(t,t_w=0)\approx A \,C_{\bf q}(t)+(1-A) \,C^{eq}_q(t),
\end{eqnarray}
with a coefficient $0<A<1$. The correlator following switch off can hence be regarded as a superposition of stationary and equilibrium correlators.

Using Eq.~\eqref{wtd_2}, we can also compute the correlator for small positive waiting times by recalling the definition of the waiting time derivative,   
\begin{equation}
C_{\bf q}(t,t_w\ll\dot\gamma^{-1})=C_{\bf q}(t,t_w=0)+\left.\frac{\partial}{\partial t_w}C_{\bf q}(t,t_w)\right|_{t_w=0}\;t_w.
\label{eq:smwai}
\end{equation} 
A comparison of our theoretically calculated correlator for finite waiting time provides a further 
test of our approximation (\ref{wtd_2}) for the waiting time derivative. 
In fact, the requirement that an approximation for the waiting time derivative generates both a sensible 
switch-off correlator (via Eq.(\ref{eq:approx})) and physically reasonable behaviour for small but finite 
waiting times would appear to be a rather demanding constraint. 

In order to calculate the MSDs as defined in the main text, we have to take the $q\to 0$ limit of  Eq.~\eqref{eq:approx}. 
This leads, without further approximation, to 
\begin{equation}
\delta r^2(t,t_w=0)\approx \delta r^2(t)+\int_0^t dt' 
\left.\frac{\partial}{\partial t_w}\delta r^2(t',t_w)\right|_{t_w=0},
\label{msd}
\end{equation}
where $\delta r^2(t)$ denotes the mean squared displacement in the NESS. Taking the limit of $q\to0$ in Eq.~\eqref{wtd_2} for the waiting time derivative, we arrive at
\begin{equation}                                                                
\delta r^2(t,t_w=0)\approx \delta r^2(t)+\int_0^t dt' \frac{\delta r^2_{eq}(t')-\delta r^2(t',t_w=0)}{\tau_\sigma},
\label{msd_schematic}                                                           
\end{equation}
where $\delta r^2_{eq}(t)$ is the MSD in the equilibrium state. Finally, the desired $\delta r^2(t,t_w=0)$ is selfconsistently expressed in terms of known MSDs. Small positive waiting times can be computed by the analog of Eq.~\eqref{eq:smwai}.

\bibliography{lit}
\bibliographystyle{apsrev4-1}